\documentclass[onecolumn,draft]{elsart3p}

\bibstyle{elsart-num}

\usepackage{amssymb}

\journal{arXiv}

\begin{document}
\begin{frontmatter}

\title{On Brownian motion in ideal gas and related principles}

\author{Yuriy E. Kuzovlev}
\ead{kuzovlev@kinetic.ac.donetsk.ua}
\address{Donetsk Institute for Physics and Technology of NASU,
ul.\,R.Luxemburg\,72\,, 83114 Donetsk, Ukraine}

%\date{\today}

\begin{abstract}
Brownian motion of particle interacting with atoms of ideal gas is
discussed as a key problem of kinetics lying at the border between
``dead'' systems like the Lorentz gas or formal constructs of
conceptual Boltzmannian kinetics and actual ``alive'' systems like
mere gas possessing scaleless (1/f) fluctuations in their kinetic
characteristics (e.g. in diffusuvity and mobility of the ``Brownian
particle'').
\end{abstract}

\begin{keyword}
Brownian motion, diffusion, molecular random walks, BBGKY hierarchy,
correlation functions, Bogolyubov functional equation, virial
relations, kinetic theory of gases

\PACS 05.20.Dd \sep 05.20.Jj \sep 05.40.Fb \sep 05.40.Jc

\end{keyword}

\end{frontmatter}

\section{Introduction (once again about derivation of
kinetics from dynamics)}

What is simpler than ideal gas? At least, when simply gas is too hard
nut to crack?

In this paper we want to consider Brownian motion of a particle
interacting with infinite gas of atoms which do not interact between
themselves. Our starting-point will be corresponding
Bogolyubov-Born-Green-Kirkwood-Yvon (BBGKY) hierarchy of equations
\cite{bog,re,bal} for $\,(n+1)$-particle distribution functions of
the ``Brownian particle'' and $\,n\,$ atoms ($\,n=0,1,2,...\,$) or
equivalent Bogolyubov functional evolution equation (BFE) for their
generating functional \cite{bog}. Our task here is not to solve these
equations but only discern them and discuss their possible solution
in terms of reasonably introduced $\,(n+1)$-particle correlation
functions, with taking in mind general invariance properties of BFE
found in \cite{last} and expressed by so-called ``virial expansion of
Brownian path probability distribution'' and other ``virial
relations''. The latter were foreseen earlier in \cite{may,mpa2} and
partly deduced in \cite{pro,mpa3,jstat,jsp} from the generalized
fluctuation-dissipation relations \cite{j123,p}.

The modesty of our task is opposed to pretence of the legend existing
among ``nutcrackers'' and stating that dilute gas (under the
``Boltzmann-Grad limit'' or, in other words, the ``low density
limit'') is so much simple object that in respect to it the BBGKY
hierarchy reduces to so-called Boltzmann hierarchy and then to the
single Boltzmann equation or some its derivative like the
Boltzmann-Lorentz equation \cite{vblls}. The legend refers to works
by Lanford on gas of hard spheres, e.g. \cite{ol0,ol} (see also
\cite{vblls} and references in \cite{bal} and \cite{vblls,ol}).
However, careful reading of work \cite{ol} shows that factually it
does not suggest a proof of the proclaimed statement. Moreover, the
author himself indicates that already short time evolution (five
times shorter than mean free-path time) reveals definite surfaces in
$\,n$-particle phase spaces ($\,n>1\,$) where $\,n$-particle
distribution functions (DF) ``do not converge to the desired
products'' of one-particle DF, thus preventing the desired proof.

The mentioned (hyper-)surfaces are $\,{\bf r}_i-{\bf r}_j=({\bf
v}_i-{\bf v}_j)\,\theta\,$ with $\,{\bf r}\,$ and $\,{\bf v}\,$ being
particle coordinates and velocities and $\,\theta\,$ arbitrary time
parameter. That are surfaces made of various pre- or post-collision
trajectories of gas particles. Hence, a correct conclusion what
suggested itself in \cite{ol} was that generally statistics of
collisions is not determined by one-particle DF, and BBGKY equations
can not be reduced to Boltzmann equation (BE), even under the ``low
density limit''.

From physical point of view, this is quite trivial conclusion.
Indeed, the ``low density limit'' in no way removes fluctuations in
density of collisions (number of pairs in pre- or post-collision
states per unit volume) as well as it does not remove fluctuations in
density of particles (number of particle per unit volume), sooner it
strengthens all them. Therefore, even if local density of collisions
was primitively thought as square function of local density of
particles, an unbiased reasoning would result in the stated
conclusion merely because of general inequality $\,\langle A*B\rangle
\neq \langle A\rangle *\langle B\rangle\,$ (with angle brackets
denoting ensemble average)!

In later attempts to derive Boltzmann hierarchy from the BBGKY one,
e.g. in \cite{vblls}, it was postulated that values of DF for an
incoming, or pre-collision, state (at $\,\theta <0\,$ with above
introduced $\,\theta \,$) and responding outgoing, or post-collision,
state (at $\,\theta >0\,$) are equal. It is good idea (although wrong
in literal quantitative sense). But elementary logics requires to
apply it at once in all places of the BBGKY hierarchy where
differential $\,n$-particle Lioville operators act on DF just
realizing collisional transitions from in- to out- states. Instead,
the authors applied their postulate in integral terms only, while in
the differential terms neglected collisions at all! Of course,
results of such arbitrariness hardly can be attributed to physics.

More likely approach to kinetic theory of dilute gas (formally, in
the Boltzmann-Grad limit) was suggested in \cite{i1} (or see
\cite{i2}). There it was emphasized that reformulation of BBGKY
hierarchy in terms of collisions of particles and ``collision
integrals'', in place of continuous interactions, is possible only if
various consecutive stages of any collision process are presented in
statistical ensemble under consideration with equal weights
(probability densities). In other words, derivative of DF in respect
to the ``inner time of collision'' $\,\theta\,$ turns to zero inside
a spatial region assigned to collision. This property never realizes
exactly, but it must be postulated in order to construct a correct
``collisional approximation'' to BBGKY hierarchy. It means \cite{i1}
that {\bf in any particular Liouville operator} the replacement
\[
-\frac {{\bf p}_i}{m}\cdot\frac {\partial }{\partial {\bf r}_i}-\frac
{{\bf p}_j}{m}\cdot\frac {\partial }{\partial {\bf r}_j}+
\Phi^{\prime}({\bf r}_i-{\bf r}_j)\cdot \left(\frac {\partial
}{\partial {\bf p}_i}-\frac {\partial }{\partial {\bf p}_j}\right)\,=
\]
\[
\,\,\,\,\,\,\,\,\,\,\,\,\,\,\,\,\,\,\,\,\,\,\,\,\,
\,\,\,\,\,\,\,\,\,\,\,\,\,\,\,\,\,\,\,\,\,\,\,\,\,\,\,
\,\,\,\,\,\,\,\,\,\,\,=\,-\frac {{\bf p}_i+{\bf
p}_j}{2m}\cdot\left(\frac {\partial }{\partial {\bf r}_i}+\frac
{\partial }{\partial {\bf r}_j}\right)+\frac {\partial }{\partial
\theta}\,\,\rightarrow\,\, -\frac {{\bf p}_i+{\bf
p}_j}{2m}\cdot\left(\frac {\partial }{\partial {\bf r}_i}+\frac
{\partial }{\partial {\bf r}_j}\right)
\]
should be made {\bf within collision} of $\,i$-th and $\,j$-th
particles, where $\,\Phi({\bf r})\,$ and $\,-\Phi^{\prime}({\bf
r})=-\partial \Phi({\bf r})/\partial {\bf r}\,$ are interaction
potential and interaction force, respectively, and the first equality
gives formal definition of the ``inner time of collision''. To be
more precise, the words ``within collision'' mean
\[
\,\,\,\,\,\,\,\,\,\,\,\,\,\,|\,{\bf r}_i-{\bf r}_j-({\bf v}_i-{\bf
v}_j)\,\theta\,|\,\lesssim \,r_0\,\,\,,\,\,\,\,\,\,|\,{\bf r}_i-{\bf
r}_j\,|\, \lesssim\,\lambda \,\,\,,
\]
with $\,r_0\,$ being characteristic interaction radius and
$\,\lambda\,$ mean free path.

Evidently, this ansatz by its nature is independent on shape of
$\,\Phi({\bf r})\,$, therefore extends to hard spheres too. It is
just what was deficient in speculations of \cite{vblls}
\footnote{This loss demonstrates that hard spheres is
treacherous model which may play a bad trick on its makers,
at least when treated in the language of DF. One can see also what
is principal error of \cite{vblls,ol0,ol} and similar
mathematical ``fantasies'': although Lebesgue measure of regions
``within collision'' tends to zero under the Boltzmann-Grad limit,
this is not a ground to cast such regions since just they determine
actual evolution of the system!}.

As the consequence, {\bf density of collisions determined by pair DF
within collision} drifts in space with velocity $\,({\bf v}_i+{\bf
v}_j)/2\,$. Since relative motion of colliding particles is included
to collision, the latter as a whole moves with this centre of mass
velocity! This is sufficient to forbid factorization of density of
collisions into any quadratic functional (e.g. product) of two
one-particle DF drifting with different velocities, $\,{\bf v}_i\,$
and $\,{\bf v}_j\,$. Hence, pair DF of colliding particles {\bf is
independent statistical characteristics} of gas evolution.

Eventually, ``collisional approximation'' to kinetics of dilute gas
results in an infinite hierarchy of kinetic equations \cite{i1,i2,p1}
for usual one-particle DF and infinitely many less usual mutually
independent $\,n$-particle DF. At $\,n>2\,$ they represent
ensemble-average densities of connected clusters of $\,n-1\,$ (real
or virtual) collisions. This new hierarchy reduces to Boltzmann
hierarchy and then to BE in case of strictly spatially uniform
statistical ensemble only (in this case, possibly, $\,\langle
A*B\rangle =\langle A\rangle *\langle B\rangle\,$ since ensemble
average can be identified with infinite-volume average) .

The word ``kinetic'' here means that interactions are represented by
a ready-made ``collision integrals'' instead of potential
$\,\Phi({\bf r})\,$. But, strictly speaking, this only gently
simplifies the theory, since full list of arguments and parameters of
$\,n$-particle DF ($\,n>1\,$) decreases by one only (namely, by the\,
$\,\theta\,$) \cite{i2}. Therefore, to derive benefit from the new
hierarchy, some its further reasonable posterization and/or
approximation is necessary. Two different approaches to this task
were considered in \cite{i1,i2} and \cite{p1,p2}.

Most principal result of \cite{i1,i2} was presence of long-living
statistical self-correlations in random (``Brownian'') motion of any
gas particle and corresponding low-frequency scaleless 1/f-type
fluctuations (``1/f-noise'') in its self-diffusivity (and mobility).
In \cite{p1,p2} this result was confirmed and seriously extended by
showing that probability density distribution of diffusive path of a
test particle possesses power-law long tails, naturally cut off at
distances of ballistic flight (see also \cite{jstat}). All this
qualitatively justifies early phenomenological theory of 1/f-noise
under realistic molecular Brownian motion \cite{bk12,bk3} and, in
turn, recently was justified in \cite{pro,mpa3,jstat,jsp} and then in
\cite{last} basing on exact invariance properties of Lioville
operator and BBGKY hierarchy.

And all this is minimum of what is neglected by the Boltzmann
hierarchy and Boltzmannian kinetics at all. Unfortunately, its modern
admirers think on pioneer level with Boltzmann, as if they are
unacquainted with Krylov's warning \cite{kr} that physical
independence of particles or events on concrete phase trajectory (in
real life) does not imply statistical independence of their images at
statistical ensemble of trajectories (``ensemble of lives''). May be,
from here the secret wish arises to equalize in rights ``dilute gas''
and ``Lorentz gas'' (non-interacting particles in a lattice of fixed
elastic scatterers).

In fact, results of \cite{i1,i2,p1,p2}, as well as
\cite{pro,mpa3,jstat,jsp,last}, say that ``dilute gas'' differs from
``Lorentz gas'' like a living matter differs from dead one. In the
first, contrary to the second, random walk of a test particle never
can be divided into statistically independent constituent parts,
regardless of their durations and total observation time. In other
words, from the point of view of time averaging, every trajectory of
this random walk remains unique at arbitrary growing spatial-temporal
scales, while in ``Lorentz gas'' almost all trajectories become
identical (use of terms ``ergodic'' or ``non-ergodic'' I leave to
mathematicians). A simple heuristic interpretation of this property
was suggested in \cite{p1} and continued in preambles of
\cite{pro,mpa3}.

Additional explanations can be found in
\cite{i1,i2,bk3} and in preambles or resumes of
\cite{jstat,p2,bk12,i3,kmg,kmg1}.

By the way, notice that the mentioned difference long ago is known as
experimental fact concerning charge carriers in semiconductor
crystals: the greater is concentration of hard immovable impurity
atoms, the stronger is damping of relative 1/f fluctuations in
carrier's diffusivity (and mobility) due to phonons (see e.g.
\cite{bk3} and references therein).

The aforesaid makes clear urgency of Brownian motion of a particle
interacting with ideal gas. Evidently, this problem mediates between
``dilute gas'' and ``Lorentz gas'', especially under the
Boltzmann-Grad limit when radius of the Brownian particle is
arbitrary small as compared with its (fixed) mean free path. We
expect that Brownian motion in ideal gas also is alive, contrary to
dead random motion among static scatterers. Let us consider this
expectation starting from \cite{last} and taking in mind experience
of \cite{i1,p1}.

\section{Functions and equations of the model}

As in \cite{last}, in the meantime we confine ourselves by Browniam
motion in equilibrium fluid (to be ideal gas here), for concreteness,
in the framework of canonical Gibbs ensemble of initial conditions
for fluid atoms. What is for the Brownian particle (BP), again
initially it is certainly located at given point $\,{\bf R}_0\,$. But
its initial momentum is random and obeys equilibrium Maxwell
distribution (with the same temperature $\,T\,$).

Let $\,\Phi({\bf r})\,$ denotes potential of interaction between BP
and atoms. Since now atoms do not interact one with another, the
relations between DF $\,F_n(t,{\bf R},{\bf r}^{(n)},{\bf P},{\bf
p}^{(n)}|{\bf R}_0;\,\nu)$ and correlation functions (CF)\,
$\,V_n(t,{\bf R},{\bf r}^{(n)},{\bf P},{\bf p}^{(n)}|{\bf
R}_0;\,\nu)\,$, with  $\,{\bf r}^{(n)}=\{{\bf r}_1...\,{\bf
r}_n\}\,$, $\,{\bf p}^{(n)}=\{{\bf p}_1...\,{\bf p}_n\}$\, and
$\,\nu\,$ standing for mean density of gas (density at infinity), the
BBGKY equations and the Bogolyubov functional evolution equation
(BFE) all strongly simplify. The BBGKY equations take form
\begin{equation}
\frac {\partial F_0}{\partial t}\,=\,-\frac {\bf P}{M}\cdot\frac
{\partial F_0}{\partial {\bf R}}\,+\,\nu\, \frac {\partial }{\partial
{\bf P}}\int_{1} \Phi^{\prime}({{\bf R}-\bf
r}_{1})\,F_{1}\,\,\,,\label{f0}
\end{equation}
\begin{equation}
\frac {\partial F_n}{\partial t}\,=\,\widehat{L}_{n}\,F_n\,+\,\nu\,
\frac {\partial }{\partial {\bf P}}\int_{n+1} \Phi^{\prime}({{\bf
R}-\bf r}_{n+1})\,F_{n+1}\,\,\,\label{fn}
\end{equation}
at $\,n>1\,$, with $\,\int_k...\,\equiv\,\int ...\,d{\bf p}_k\,d{\bf
r}_k\,$,\,
\begin{equation}
\widehat{L}_n\,\equiv\,-\frac {\bf P}{M}\cdot\frac {\partial
}{\partial {\bf R}}\,-\sum_{j\,=1}^n \frac {{\bf p}_j}{m}\cdot\frac
{\partial }{\partial {\bf r}_j}\,+\sum_{j\,=1}^n\Phi^{\prime }({\bf
R}-{\bf r}_{j})\cdot \left(\frac {\partial }{\partial {\bf P}}-\frac
{\partial }{\partial {\bf p}_j}\right )\,\,\,,\label{lon}
\end{equation}
and obvious initial conditions:
\begin{equation}
\begin{array}{l}
F_n(t=0,...\,)\,=\, \delta({\bf R}-{\bf R}_0)\,G_M({\bf
P})\prod_{j\,=1}^n G_m({\bf p}_j)\,E({\bf r}_j)\,\,\,,\label{ic}\\
G_m({\bf p})\,\equiv\,(2\pi Tm)^{-\,3/2}\exp{(-{\bf
p}^2/2Tm)}\,\,\,,\,\,\,\,\,\,E({\bf r})\,\equiv\,\exp{[-\,\Phi({\bf
r})/T\,]}
\end{array}
\end{equation}
The connection between generating functional of DF,
\[
\mathcal{F}\{t,{\bf R},{\bf P},\psi\,|{\bf R}_0;\nu\}=F_0(t,{\bf
R},{\bf P}|{\bf R}_0;\,\nu)+\sum_{n\,=1}^{\infty } \frac
{\nu^n}{n!}\int_1 ...\int_n F_n(t,{\bf R},{\bf r}^{(n)},{\bf P},{\bf
p}^{(n)}|{\bf R}_0;\nu)\prod_{j\,=1}^n \psi({\bf r}_j,{\bf
p}_j)\,\,\,,
\]
and quite similarly defined generating functional of CF \cite{last},
$\,\mathcal{V}\{t,{\bf R},{\bf P},\psi\,|{\bf R}_0;\nu\}\,$, becomes
\begin{equation}
\begin{array}{l}
\mathcal{F}\{t,{\bf R},{\bf P},\,\psi\,|{\bf R}_0;\,\nu\}\,\,=\,
\,\exp{\left[\,\nu\int G_m({\bf p})\,E({\bf r}-{\bf R})\,\psi({\bf
r},\bf p)\,d{\bf p}\,d{\bf r}\,\right]}\,\,\mathcal{V}\{t,{\bf
R},{\bf P},\,\psi\,|{\bf
R}_0;\,\nu\}\,=\,\\
\,\,\,\,\,\,\,\,\,\,\,\,\,\,\,\,\,\,\,\,=\,\exp{\left[\,\nu\int
E({\bf r}-{\bf R})\,\phi({\bf r})\,d{\bf
r}\,\right]}\,\,\mathcal{V}\{t,{\bf R},{\bf P},\,\psi\,|{\bf
R}_0;\,\nu\}\,\,\,,\label{vf}
\end{array}
\end{equation}
where\, $\,\phi({\bf r})\,\equiv\, \int \psi({\bf r},{\bf
p})\,G_m({\bf p})\,d{\bf p}\,$. Recall that in fact this is
definition of CF. According to it,
\[
F_0(t,{\bf R},{\bf P}| {\bf R}_0;\nu)=V_0(t,{\bf R},{\bf P}| {\bf
R}_0;\nu)\,\,\,,
\]
\begin{equation}
\begin{array}{l}
F_1(t,{\bf R},{\bf r}_1,{\bf P},{\bf p}_1|{\bf R}_0;\nu)=V_0(t,{\bf
R},{\bf P}|{\bf R}_0;\nu)\,E({\bf r}_1-{\bf R})\,G_m({\bf p}_1)+
V_1(t,{\bf R},{\bf r}_1,{\bf P},{\bf p}_1|{\bf R}_0;\nu)\,\,\,,
\label{cf1}
\end{array}
\end{equation}
and so on. The BFE, that is compact functional form of BBGKY
hierarchy, now reads
\begin{equation}
\begin{array}{l}
\frac {\partial \mathcal{F}}{\partial t}\,+\,\frac {\bf
P}{M}\cdot\frac {\partial \mathcal{F}}{\partial {\bf R}}\,=\,-\int
\psi(x)\, \frac {{\bf p}}{m}\cdot\frac {\partial }{\partial {\bf
r}}\,\frac {\delta \mathcal{F}}{\delta \psi(x)}\,\,+\,\label{fe}\int
[\,1+\psi(x)\,]\,\,\Phi^{\prime }({\bf R}-{\bf r})\cdot \left(\frac
{\partial }{\partial {\bf P}}-\frac {\partial }{\partial {\bf
p}}\right )\frac {\delta \mathcal{F}}{\delta \psi(x)}\,\label{fe}
\end{array}
\end{equation}
with\, $\,x\,=\{{\bf r},{\bf p}\}\,$\, and\, $\,\int ...\,=\int
...\,dx\,=\int ...\,d{\bf p}\,d{\bf r}\,$,\, thus turning into
first-order differential equation in respect to $\,\psi(x)\,$.
Substitution of (\ref{vf}) to (\ref{fe}) yields equivalent BFE in
terms of CF:
\begin{equation}
\begin{array}{l}
\frac {\partial \mathcal{V}}{\partial t}\,+\,\frac {\bf
P}{M}\cdot\frac {\partial \mathcal{V}}{\partial {\bf R}}\,=\,-\int
\psi(x)\, \frac {{\bf p}}{m}\cdot\frac {\partial }{\partial {\bf
r}}\,\frac {\delta \mathcal{V}}{\delta \psi(x)}\,\,+\,\int
[\,1+\psi(x)\,]\,\,\Phi^{\prime }({\bf R}-{\bf r})\cdot \left(\frac
{\partial }{\partial {\bf P}}-\frac {\partial }{\partial {\bf
p}}\right )\frac {\delta \mathcal{V}}{\delta \psi(x)}\,\,+\,\label{fev}\\
\,\,\,\,\,\,\,\,\,\,\,\,\,\,\,\,\,\,\,\,\,\,\,\,\,
\,\,\,\,\,\,\,\,\,\,\,\,\,\,\,\,\,\,\,\,\,\,\,\,\,\,
+\,\nu\,T\left[\int G_m({\bf p})\,E^{\,\prime}({\bf r}-{\bf
R})\,\,\psi(x)\,\right]\cdot\left(\frac {{\bf P}}{MT}+\frac {\partial
}{\partial {\bf P}}\right )\,\mathcal{V}\,\,\,,
\end{array}
\end{equation}
with\, $\,E^{\,\prime}({\bf r})=dE({\bf r})/d{\bf r}=-\Phi^{\prime
}({\bf r})E({\bf r})/T\,$.\,

Variational differentiations of (\ref{fev}) return us to BBGKY
equations rewritten in terms of CF:
\begin{equation}
\frac {\partial V_0}{\partial t}\,=\,-\frac {\bf P}{M}\cdot\frac
{\partial V_0}{\partial {\bf R}}\,+\,\nu\, \frac {\partial }{\partial
{\bf P}}\int_{1} \Phi^{\prime}({{\bf R}-\bf
r}_{1})\,V_{1}\,\,\,,\label{v0}
\end{equation}
\begin{equation}
\frac {\partial V_1}{\partial t}\,=\,\widehat{L}_{1}\,V_1\,+\,\nu\,
\frac {\partial }{\partial {\bf P}}\int_{2} \Phi^{\prime}({{\bf
R}-\bf r}_{2})\,V_{2}\,+\,\,T\,G_m({\bf p}_1)\,E^{\,\prime}({\bf
r}_1-{\bf R})\left(\frac {{\bf P}}{MT}+\frac {\partial }{\partial
{\bf P}}\right )V_0\,\,\,,\label{v1}
\end{equation}
\begin{equation}
\begin{array}{l}
\frac {\partial V_n}{\partial t}\,=\,\widehat{L}_{n}\,V_n\,+\,\nu\,
\frac {\partial }{\partial {\bf P}}\int_{n+1} \Phi^{\prime}({{\bf
R}-\bf r}_{n+1})\,V_{n+1}\,+\,\,\,\label{vn}\\
\,\,\,\,\,\,\,\,\,\,\,\,\,\,\,\,\,\,+\,\,\,T\,\sum_{j\,=1}^n G_m({\bf
p}_j)\,E^{\,\prime}({\bf r}_j-{\bf R})\cdot\left(\frac {{\bf
P}}{MT}+\frac {\partial }{\partial {\bf P}}\right )
V_{n-\,1}(...\,{\bf r}_{j-1},{\bf r}_{j\,+1}...\,{\bf p}_{j-1},{\bf
p}_{j\,+1}\,...)\,\,
\end{array}
\end{equation}
Initial conditions and boundary conditions to these equations and
(\ref{fev}) are very simple:
\begin{equation}
\begin{array}{l}
V_n(t=0,...\,)\,=\,\delta_{n,\,0}\,\,\delta({\bf R}-{\bf
R}_0)\,G_M({\bf P}) \,\,\,,\,\,\,\,\,\,\mathcal{V}\{t=0,\,{\bf
R},{\bf P},\,\psi\,|{\bf R}_0;\nu\}\,=\,\delta({\bf R}-{\bf
R}_0)\,G_M({\bf P})\,\,\,\label{icv}
\end{array}
\end{equation}
(I have to detect misprint in similar formula (13) in \cite{last}:
there factor $\,G_M({\bf P})\,$ was missed!),
\begin{equation}
\begin{array}{l}
V_{n>\,0}(\,...\,{\bf r}_k\,...\,)\,\rightarrow
\,0\,\,\,\,\,\,\,\,\texttt{at}\,\,\,\,\,\,\,{\bf r}_k-{\bf
R}\,\rightarrow\, \infty\,\,\label{bcv}
\end{array}
\end{equation}

At last, consider the invariance group found in \cite{last} (a group
of such transformations of arguments of generating functional
determined by (\ref{fev})-(\ref{bcv}) which do not change value of
the functional). In case of ideal gas it simplifies to
\begin{equation}
\begin{array}{l}
\mathcal{V}\{t,{\bf R},{\bf P},\,\sigma +\psi\,|\,{\bf
R}_0;\,\nu\}\,=\,\mathcal{V}\{t,{\bf R},{\bf P},\,
\psi/(1+\sigma)\,|\,{\bf R}_0;\,(1+\sigma)\,\nu\}\,\,\,,\label{id}
\end{array}
\end{equation}
where $\,\sigma({\bf r},{\bf p})=\sigma =\,$\,const\, is arbitrary
constant from interval $\,-1<\sigma <\infty\,$. This functional
identity implies exact ``virial expansions'' of CF:
\begin{eqnarray}
\begin{array}{l}
V_n(t,{\bf R},{\bf r}^{(n)},{\bf
P},{\bf p}^{(n)}|{\bf R}_0;\,\nu+\sigma \nu )\,=\,\\
\,\,\,\,\,\,\,\,\,\,\,\,\, =\,V_n(t,{\bf R},{\bf r}^{(n)},{\bf
P},{\bf p}^{(n)}|{\bf R}_0;\,\nu)\,+\sum_{k\,=\,1}^\infty \frac
{(\nu\sigma)^{\,k}}{k!}\int_{n+1} ...\int_{n+k} V_{n+k}(t,{\bf
R},{\bf r}^{(n+k)},{\bf P},{\bf p}^{(n+k)}|{\bf
R}_0;\,\nu)\,\,\label{vexpn}
\end{array}
\end{eqnarray}
Their infinitesimal form yields exact ``virial relations''
\begin{eqnarray}
\begin{array}{l}
\frac {\partial V_n(t,{\bf R},{\bf r}^{(n)},{\bf P},{\bf
p}^{(n)}|{\bf R}_0;\,\nu)}{\partial \nu}\,=\,\int_{n+1}
V_{n+1}(t,{\bf R},{\bf r}^{(n+1)},{\bf P},{\bf p}^{(n+1)}|{\bf
R}_0;\,\nu) \,\,\label{inf}
\end{array}
\end{eqnarray}
At $\,n=0\,$, in particular, we obtain connection between density
derivative of the probability distribution of path, $\,\Delta{\bf
R}={\bf R}-{\bf R}_0\,$, of the Brownian particle and integrated pair
CF:
\begin{eqnarray}
\begin{array}{l}
\frac {\partial V_0(t,{\bf R},{\bf P}|{\bf R}_0;\,\nu)}{\partial
\nu}\,=\,\int V_1(t,{\bf R},{\bf r},{\bf P},{\bf p}|{\bf
R}_0;\,\nu)\,\,d{\bf r}\,d{\bf p}\,\,\,,\label{inf0}
\end{array}
\end{eqnarray}
where, of course, both sides depend on current position of BP,
$\,{\bf R}\,$, and it starting position, $\,{\bf R}_0\,$, through
their difference $\,\Delta{\bf R}={\bf R}-{\bf R}_0\,$ only.

Relations like (\ref{inf0}), as combined with identity (\ref{cf1})
and trivial positivity of DF $\,F_1\,$, lead to principal consequence
\cite{last,pro,mpa3,jstat,jsp} mentioned in Introduction: asymptotic
of BP's path distribution,
\begin{equation}
\begin{array}{l}
V_0(t,\Delta{\bf R};\,\nu)\,=\,\int V_0(t,{\bf R},{\bf P}|{\bf
R}_0;\,\nu)\,d{\bf P}\,\,\,,\label{ppd}
\end{array}
\end{equation}
at $\,t\gg\tau\,$, with $\,\tau\,$ being mean free-path time of BP,
as a function of $\,\Delta{\bf R}\,$ has power-law long tails lasting
up to $\,|\Delta{\bf R}|\sim t\sqrt{T/M}\,$.

Such statement strikingly contradicts Boltzmannian kinetics which
always gravitates towards the ``law of large numbers'' and Gaussian
asymptotic characterized by short exponential tails. To avoid new
repeating myself, I ask dear reader to search for proper comments in
\cite{mpa3,jstat,jsp,i1,i2,p1}. But one not popular truism deserves
repetition: in statistical physics, unlike probability theory, one should
not rely on the ``law of large numbers'' since practically independent
events or quantities may be nevertheless statistically dependent. It
would be a good thing to understand in detail how all this realize in
case of Brownian motion in ideal gas.

Notice that due to simplicity of BBGKY equations
(\ref{f0})-(\ref{fn}) or (\ref{v0})-(\ref{vn}), in comparison with
general case \cite{last,i1,p1}, one can easy verify virial relations
(\ref{inf}) by deriving them directly from (\ref{v0})-(\ref{vn}).

\section{Relative coordinates and characteristic function}
Eventually, we would like to obtain the probability distribution
(\ref{ppd}) of BP's path or its characteristic function, that is
Fourier transform
\begin{equation}
\begin{array}{l}
V_0(t,i{\bf k};\,\nu)\,\equiv\,\int \exp{(i{\bf k}\cdot\Delta{\bf
R})}\,V_0(t,\Delta{\bf R};\,\nu)\,d\Delta{\bf R}\,\,\label{nf0}
\end{array}
\end{equation}
Therefore, first, let us consider all the DF and CF as
functions of $\,\Delta{\bf R}={\bf R}-{\bf R}_0\,$ and relative
distances\, $\,{\bf \rho}_j={\bf r}_j-{\bf R}\,$\,. Such change
of spatial variables implies operator changes
\[
\frac {\partial }{\partial {\bf R}}\,\rightarrow\,\frac {\partial
}{\partial \Delta{\bf R}}-\sum_j\frac {\partial }{\partial
\rho_j}\,\,\,,\,\,\,\,\,\,\frac {\partial }{\partial {\bf
r}_j}\,\rightarrow\,\frac {\partial }{\partial \rho_j}
\]
everywhere in (\ref{f0})-(\ref{fn}), (\ref{v0})-(\ref{vn}).
Second, make Fourier transform in respect to
$\,\Delta{\bf R}\,$ and go to functions
\begin{equation}
\begin{array}{l}
V_n(t,i{\bf k},{\bf \rho}^{(n)},{\bf P},{\bf
p}^{(n)};\,\nu)\,\equiv\,\int \exp{[i{\bf k}\cdot({\bf R}-{\bf
R}_0)]}\,\,V_n(t,{\bf R},{\bf R}+{\bf \rho}^{(n)},{\bf P},{\bf
p}^{(n)}|{\bf R}_0;\,\nu)\,\,d{\bf R}\,\label{nf}
\end{array}
\end{equation}
Third, introduce velocities \, $\,{\bf V}\equiv {\bf P}/M\,$\, and\,
$\,{\bf v}_j\equiv {\bf p}_j/m\,$\, of BP and gas atoms, and new
operator
\begin{equation}
\,\,\,\widehat{\Lambda}(j)\,\,\equiv\,-\,({\bf v}_j-{\bf
V})\cdot\frac {\partial}{\partial
\rho_j}\,+\,\Phi^{\prime}(\rho_j)\cdot\left(\frac {\partial}{\partial
{\bf p}_j}-\frac {\partial}{\partial {\bf P}}\right)\,\,\label{nlo}
\end{equation}
Then BBGKY equations (\ref{v0})-(\ref{vn}) take the form
\begin{equation}
\frac {\partial V_0}{\partial t}\,=\,i({\bf k}\cdot{\bf
V})\,V_0\,-\,\nu\, \frac {\partial }{\partial {\bf P}}\int_{1}
\Phi^{\prime}(\rho_{\,1})\,V_{1}\,\,\,,\label{v01}
\end{equation}
\begin{equation}
\frac {\partial V_1}{\partial t}\,=\,i({\bf k}\cdot{\bf
V})\,V_1\,+\,\widehat{\Lambda}(1)\,V_1\,-\,\nu\, \frac {\partial
}{\partial {\bf P}}\int_{2}
\Phi^{\prime}(\rho_{\,2})\,V_{2}\,+\,\,G_m({\bf
p}_1)\,E^{\,\prime}(\rho_1)\left({\bf V}+T\,\frac {\partial
}{\partial {\bf P}}\right )V_0\,\,\,,\label{v11}
\end{equation}
\begin{equation}
\begin{array}{l}
\frac {\partial V_n}{\partial t}\,=\,i({\bf k}\cdot{\bf
V})\,V_n\,+\,\sum_{j\,=1}^n\widehat{\Lambda}(j)\,V_n\,-\,\nu\, \frac
{\partial }{\partial {\bf P}}\int_{n+1} \Phi^{\prime}(\rho_{\,n+1})\,
V_{n+1}\,+\,\,\,\label{vn1}\\\,\,\,\,\,\,\,\,\,\,\,\,\,\,\,\,\,\,
\,\,\,\,\,\,\,\,\,\,\,\,\,\,\,\,\,\,+\,\,\,\sum_{j\,=1}^n G_m({\bf
p}_j)\,E^{\,\prime}(\rho_j)\cdot\left({\bf V}+T\,\frac {\partial
}{\partial {\bf P}}\right )
V_{n-\,1}(...\,\rho_{j-1},\rho_{j\,+1}...\,{\bf p}_{j-1},{\bf
p}_{j\,+1}\,...)\,\,
\end{array}
\end{equation}
with initial and boundary conditions as follow,
\begin{equation}
\begin{array}{l}
V_n(t=0,...\,)\,=\,\delta_{n,\,0}\,G_M({\bf P})
\,\,\,,\,\,\,\,\,\,V_{n>\,0}(\,...\,\rho_k\,...\,)\,\rightarrow
\,0\,\,\,\,\,\,\,\,\texttt{at}\,\,\,\,\,\,\,\rho_k\,\rightarrow\,
\infty\,\,\,,\label{ibc}
\end{array}
\end{equation}
and BP's path distribution (\ref{ppd}) under interest presented by
\begin{equation}
\begin{array}{l}
V_0(t,\Delta{\bf R};\,\nu)\,=\,\int \exp{(-i{\bf k}\cdot\Delta{\bf
R})}\int V_0(t,i{\bf k},{\bf P};\,\nu)\,d{\bf P}\,\,\frac {d{\bf
k}}{(2\pi)^{d}}\,\,\,,\label{nppd}
\end{array}
\end{equation}
where $\,d\,$ is space dimension ($\,d=3\,$ by default).

Taking into account quite obvious identity\, $\,({\bf V}+T\,\partial
/\partial {\bf P} )\,G_M({\bf P})=0\,$\, it is easy to see that at\,
$\,{\bf k}=0\,$ the system (\ref{v01})-(\ref{vn1}) always stays in
its initial state:
\begin{equation}
\begin{array}{l}
V_0(t,i{\bf k}=0,{\bf P};\,\nu)\,=\,G_M({\bf P})\,\,\,, \label{k0}\\
V_n(t,i{\bf k}=0,{\bf \rho}^{(n)},{\bf P},{\bf
p}^{(n)};\,\nu)\,=\,\int V_n(t,{\bf R},{\bf R}+\rho^{(n)},{\bf
P},{\bf p}^{(n)}|{\bf R}_0;\,\nu)\,\,d{\bf
R}\,=\,0\,\,\,\,\,\,\,\,\,\,\,\,\,\,\,\,\,\,(n>0)\,
\end{array}
\end{equation}
First of these expresses mere normalization of BP's path distribution
and besides says that (unconditional) probability distribution of
BP's velocity in equilibrium gas always stays equilibrium.

Collecting all CF $\,V_n(t,i{\bf k},{\bf \rho}^{(n)},{\bf P},{\bf
p}^{(n)};\,\nu)\,$ into generating functional
\[
\mathcal{V}\{t,i{\bf k},{\bf P},\psi\,;\nu\}\,\equiv\,V_0(t,i{\bf
k},{\bf P};\,\nu)+\sum_{n\,=1}^{\infty } \frac {\nu^n}{n!}\int_1
...\int_n V_n(t,i{\bf k},\rho^{(n)},{\bf P},{\bf
p}^{(n)};\nu)\prod_{j\,=1}^n \psi(\rho_j,{\bf p}_j)\,\,\,,
\]
one may replace all equations (\ref{v01})-(\ref{vn1}) by single
functional equation:
\begin{equation}
\begin{array}{l}
\frac {\partial \mathcal{V}}{\partial t}\,=\,\widehat{\pounds}\,
\mathcal{V}\,\,\label{fev1}
\end{array}
\end{equation}
with evolution operator
\begin{eqnarray}
\widehat{\pounds}\,=\,\,i({\bf k}\cdot{\bf V})\,+\int
\psi(x)\left[\,({\bf V}-{\bf v})\cdot\frac {\partial}{\partial
\rho}\,+\,\Phi^{\prime}(\rho)\cdot\left(\frac {\partial}{\partial
{\bf p}}-\frac {\partial}{\partial {\bf P}}\right)\,\right]\frac
{\delta }{\delta
\psi(x)}\,-\,\label{mop}\\
-\,\frac {\partial }{\partial {\bf P}} \int
\Phi^{\prime}(\rho)\,\frac {\delta }{\delta
\psi(x)}\,+\,\nu\left[\int G_m({\bf
p})\,E^{\,\prime}(\rho)\,\psi(x)\,\right]\cdot\left({\bf V}+T\frac
{\partial }{\partial {\bf P}}\right)\,\,\,\,\,\nonumber
\end{eqnarray}
and initial condition
\begin{equation}
\begin{array}{l}
\mathcal{V}\{t=0,i{\bf k},{\bf P},\psi\,;\nu\}\,=\,G_M({\bf
P})\,\,\label{icv1}
\end{array}
\end{equation}
In essence, of course, this is full equivalent of (\ref{fev})
following from (\ref{fev}) under change $\,\psi({\bf r},{\bf
p})\rightarrow\psi({\bf R}+\rho,{\bf p})\,$ and correspondingly
\[
\frac {\partial }{\partial {\bf R}}\rightarrow \frac {\partial
}{\partial {\bf R}}+\int d{\bf p}\int d\rho\,\,\frac {\partial
\psi(\rho,{\bf p})}{\partial \rho}\,\frac {\delta }{\delta
\psi(\rho,{\bf p})}
\]

\section{Boson representation and path integrals}
{\bf 1}.\, According to formulas of previous section, direct formal
exact solution of BBGKY equations (\ref{v01})-(\ref{vn1}) in respect
to the characteristic function (\ref{nf0}) of BP's path distribution
can be written as
\begin{equation}
\begin{array}{l}
V_0(t,i{\bf k};\,\nu)\,=\,\int \mathcal{V}\{t,i{\bf k},{\bf
P},\psi=0\,;\nu\}\,\,d{\bf P}\,=\,\left[\int d{\bf
P}\,\,e^{\,t\widehat{\pounds}}\,\,G_M({\bf
P})\,\right]_{\psi\,=\,0}\,\,\,,\label{fa}
\end{array}
\end{equation}
where the evolution operator $\,\widehat{\pounds}\,$ represents, in
respect to BP's momentum, a linear combination of two vector
operators,\, $\,{\bf V}+T\,\partial /\partial {\bf P}\,$ and
$\,-T\,\partial /\partial {\bf P}\,$ (factor $\,{\bf V}\,$ in first
row of (\ref{mop}) is their sum). Let us normalize them so that their
components commute one with another exactly as boson birth and
annihilation operators:
\begin{equation}
\begin{array}{l}
{\bf A}^\dag\,\equiv\,-\,\sqrt{TM}\,\frac {\partial }{\partial {\bf
P}}\,\,\,\,,\,\,\,\,\,\,{\bf A}\,\equiv\,\sqrt{\frac MT}\,\left({\bf
V}+T\frac {\partial }{\partial {\bf
P}}\right)\,\,\,,\,\,\,\,\,\,A_{\alpha}A^\dag_{\beta}
-A^\dag_{\beta}A_{\alpha}\,=\,\delta_{\alpha\beta}\,\label{aa}
\end{array}
\end{equation}
Then notice that\, $\,{\bf A}\,G_M({\bf P})=0\,$\, and\, $\,\int
d{\bf P}\,{\bf A}^\dag\,...\,=0\,$\,. Therefore the Maxwell
distribution,\, $\,G_M({\bf P})=0\,$,\, can be treated as ket-vector
of ground state of ``quantum harmonic oscillators'' represented by
$\,{\bf A}\,$ and $\,{\bf A}^\dag\,$, while integration over
momentum, $\,\int d{\bf P}\,...\,$\,, acts as multiplication by
bra-vector of the ground state.

Besides, notice or recall that operators of multiplication by
$\,\psi(x)\,$ and differentiation $\,\delta/\delta\psi(x)\,$
also behave as boson birth and annihilation operators, and we can write
\begin{equation}
\begin{array}{l}
a^\dag(x)\,\equiv\,c(x)\,\psi(x)\,\,\,\,,\,\,\,\,\,\,a(x)\,\equiv\,
c^{-1}(x)\,\frac {\delta }{\delta \psi(x)}\,\,\,,
\,\,\,\,\,\,\,\,\,a(x)a^\dag(y)-a^\dag(y)a(x)\,=
\,\delta(x-y)\,\,\,,\label{ax}
\end{array}
\end{equation}
with arbitrary fixed $\,c(x)\neq 0\,$.  At that, unit from the right
of them and putting on $\,\psi =0\,$ after their action again represent
ket- and bra-vectors of vacuum state, respectively.
A proper choice of $\,c(x)\,$\, is
\[
c(x)\,=\,\sqrt{\nu\,G_m({\bf p})\,E(\rho)}
\]
For further let us introduce also\,\,
$\,v_0\,\equiv\,\sqrt{T/m}\,\,\,,\,\,\,\,\,\,u_0\,\equiv\,\sqrt{T/M}\,$.

Then after some algebra formulas (\ref{mop}) and (\ref{fa}) can be
rewritten as
\begin{equation}
\begin{array}{l}
V_0(t,i{\bf k};\,\nu)\,=\,\langle
0|\,\,e^{\,t\widehat{\pounds}}\,\,|0\rangle \,\,\,,\label{fa1}\\
\widehat{\pounds}\,=\,\widehat{\pounds}_1\,+
\,\widehat{\pounds}_2\,+\,\widehat{\pounds}_3\,\,\,,
\end{array}
\end{equation}
where $\,|0\rangle\,$ is common ground state (``vacuum state''), and
three parts of the evolution operator are linear, quadratic and cubic
forms of the birth and annihilation operators, respectively:
\begin{equation}
\begin{array}{l}
\widehat{\pounds}_1\,=\,\,i({\bf k}\cdot({\bf A}+{\bf
A}^\dag))\,u_0\,\,\,,\,\label{mop1}
\end{array}
\end{equation}
\begin{equation}
\begin{array}{l}
\widehat{\pounds}_2\,=\,\,\int a^\dag(x)\left[-{\bf v}\cdot\frac
{\partial}{\partial \rho}\,+\,\Phi^{\prime}(\rho)\cdot\frac
{\partial}{\partial {\bf p}}\,\right]a(x)\,+\,u_0\int c(x)\,\frac
{\Phi^{\prime}(\rho)}{T}\cdot[\,a(x){\bf A}^\dag\,-\,a^\dag(x) {\bf
A}\,]\,\,\,,\label{mop2}
\end{array}
\end{equation}
\begin{equation}
\begin{array}{l}
\widehat{\pounds}_3\,=\,u_0\int a^\dag(x)\left[\,({\bf A}^\dag+{\bf
A})\cdot\frac {\partial}{\partial \rho}\,+\,({\bf A}^\dag-{\bf
A})\cdot\frac {\Phi^{\prime}(\rho)}{2T}\,\right]a(x)\,\label{mop3}
\end{array}
\end{equation}
In such way calculation of exact characteristic function of BP's path
transforms to calculation of the ``vacuum-vacuum transition``
amplitude,\,\, $\,\langle
0|\,\,e^{\,-\,it\widehat{\mathcal{H}}}\,\,|0\rangle\,$,\, for excited
system of interacting bosons, or quantum oscillators, with cubic
Hamiltonian $\,\widehat{\mathcal{H}}=i\widehat{\pounds}\,$\,.\,\, Due
to the boundary conditions (\ref{bcv}),(\ref{ibc}) one can think
that\, $\,a^\dag(x)a(x)\rightarrow 0\,$\, at\,
$\,\rho\rightarrow\infty\,$\, in those sense that far enough
oscillators almost surely stay in their ground states. Therefore
 $\,\widehat{\mathcal{H}}\,$ can be treated as Hermitian
operator.

{\bf 2}.\, On this way, next possible step is use of so-called
holomorphic form of path integrals (see e.g. \cite{pf,sf,lf}).
According to it, we can replace (\ref{fa1}) by path integral
\begin{equation}
\begin{array}{l}
V_0(t,i{\bf k};\,\nu)\,=\,\langle
0|\,\,e^{\,t\widehat{\pounds}}\,\,|0\rangle\,=\, \int
\exp{\left\{\int^t_0 \left[ \frac 12\,
(\,\dot{\mathcal{A}}^*\mathcal{A}\,-\,\mathcal{A}^*\dot{\mathcal{A}}\,)
\,+\,\widehat{\pounds}(\mathcal{A}^*,\mathcal{A})
\right]d\xi\right\}}\,\prod_{\xi} \frac
{d\mathcal{A}^*d\mathcal{A}}{2\pi i}\,\,\label{fa2}
\end{array}
\end{equation}
supplemented with edge conditions\, $\,\mathcal{A}^*(\xi=t)=0\,$,\,
$\,\mathcal{A}(\xi=0)=0\,$\,, where
$\,\mathcal{A}^*=\mathcal{A}^*(\xi)= \{a^*(x,\xi),{\bf A}^*(\xi)\}
\,$ and $\,\mathcal{A}=\mathcal{A}(\xi) =\{a(x,\xi),{\bf A}(\xi)\}\,$
are holomorphic images of all above introduced birth and annihilation
operators in the form of complex-valued time functions, and  the dot
means derivative in respect to time $\,\xi\,$. Integrating
(\ref{fa2}) first over variables $\,a^*(x,\xi)\,$, $\,a(x,\xi)\,$
deputed by $\,a^\dag(x)\,$ and $\,a(x)\,$ one obtains
\begin{equation}
\begin{array}{l}
V_0(t,i{\bf k};\,\nu)\,=\,\int \, \exp{\left\{\int^t_0 \left[ \frac 12\,
(\dot{{\bf A}}^*\cdot{\bf A}-{\bf A}^*\cdot\dot{{\bf A}}) \,+\,i({\bf
k}\cdot({\bf A}+{\bf A}^*))\,u_0\,
\right]d\xi\right\}}\,\times\,\label{fa3}\\
\,\,\,\,\,\,\,\,\,\,\,\,\,\,\,\,\,\,\,\,\,\,\,\,\,\,\,\,\,\,\,\,
\times\,\,\exp{\left[-\int_{t>\,\xi_1>\,\xi_2>\,0} A_{\alpha}^*(\xi_1)
\,\,\mathbb{G}_{\alpha\beta}\{\xi_1,\xi_2,\,{\bf A}^*,{\bf A}\}\,\,
A_{\beta}(\xi_2)\,\,d\xi_2\,d\xi_1\right]}\,\,\prod_{\xi} \frac {d{\bf A}^*d{\bf
A}}{(2\pi i)^d}\,\,\,\,,
\end{array}
\end{equation}
with edge conditions\, $\,{\bf A}^*(t)=0\,$,\, $\,{\bf A}(0)=0\,$\,,
where repeated indices imply summation, the kernel\,
$\,\mathbb{G}_{\alpha,\,\beta}\,$\, is functional of $\,{\bf
A}^*(\xi)=0\,$ and $\,{\bf A}(\xi)=0\,$ expressed by
\begin{equation}
\begin{array}{l}
\mathbb{G}_{\alpha\beta}\{\xi_1,\xi_2,\,{\bf A}^*,{\bf A}\}\,=\,
\frac {u_0^2}{T^2}\int c(x)\,\Phi^{\prime}_{\alpha}(\rho)\,\,
\overleftarrow\exp{\left[ \int^{\xi_1}_{\xi_2} \widehat\Lambda({\bf
A}^*(\xi),{\bf
A}(\xi))\,d\xi\right]}\,\,\Phi_{\beta}^{\prime}(\rho)\,c(x)
\,\,\,,\label{kern}
\end{array}
\end{equation}
and\, $\,\widehat\Lambda({\bf A}^*,{\bf A})\,$\, is sum of evolution
operators from (\ref{mop2}) and (\ref{mop3}),
\begin{equation}
\begin{array}{l}
\widehat\Lambda({\bf A}^*,{\bf A})\,=\, -\,{\bf v}\cdot\frac
{\partial}{\partial \rho}\,+\,\Phi^{\prime}(\rho)\cdot\frac
{\partial}{\partial {\bf p}}\,+\,\,u_0\left[({\bf A}^* +{\bf
A})\cdot\frac {\partial}{\partial \rho}\,+\,({\bf A}^*-{\bf
A})\cdot\frac {\Phi^{\prime}(\rho)}{2T}\right]\,\label{L}
\end{array}
\end{equation}
(clearly, this is image of the Liouville operator (\ref{nlo})). Here
and below $\,\overleftarrow\exp\,$ designates chronologically ordered
exponential and, as before, $\,\int\,...\,=\int\int\,...\,d{\bf
p}\,d\rho\,$. With use of obvious identities
\[
\begin{array}{l}
\widehat\Lambda({\bf A}^*,{\bf A})\,c(x)\,=\, -\,\frac
{u_0\,\Phi^{\prime}(\rho)\cdot{\bf A}\,c(x)}{T}\,\,\,,\,\,\,\,\,\,
\widehat\Lambda^{\,\top}({\bf A}^*,{\bf A})\,c(x)\,=\, \frac
{u_0\,\Phi^{\prime}(\rho)\cdot{\bf A}^*\,c(x)}{T}\,\,
\end{array}
\]
(where $\,\top\,$ symbolizes transposition) one can  transform
(\ref{fa3}) into
\begin{equation}
\begin{array}{l}
V_0(t,i{\bf k};\,\nu)\,=\,\int \, \exp{\left\{\int^t_0 \left[ \frac
12\, (\dot{{\bf A}}^*\cdot{\bf A}-{\bf A}^*\cdot\dot{{\bf A}})
\,+\,i({\bf k}\cdot({\bf A}+{\bf A}^*))\,u_0\,
\right]d\xi\right\}}\,\times\,\label{fa4}\\\,\,\,\,\,\,\,\,\,\,\,\,\,\,\,
\,\,\,\,\,\,\,\,\,\,\,\,\,\,\,\,\,\,\times\,\,\exp{\left\{\int c(x)
\left(\overleftarrow\exp{\left[ \int^t_0 \widehat\Lambda({\bf
A}^*,{\bf A})\,d\xi\right]}-1\right)c(x)\right\}} \,\,\prod_{\xi}
\frac {d{\bf A}^*d{\bf A}}{(2\pi i)^d}\,
\end{array}
\end{equation}

Exact analogue of formula (\ref{fa4}) was obtained in \cite{sr}, in
slightly different notation, by means of so-called ``stochastic
representation of deterministic interactions'' (see also references
from \cite{sr}).

Integration of (\ref{fa2}) at first over $\,{\bf A}\,$ and $\,{\bf
A}^*\,$ results in another kind of holomorphic path integrals:
\begin{equation}
\begin{array}{l}
V_0(t,i{\bf k};\,\nu)\,=\,\int \exp{\left\{\int^t_0 \int\left[ \frac
12\, (\,\dot{a}^*a\,-\,a^*\dot{a}\,)\,+\,a^*\,\widehat\Lambda_0\, a
\,\right]dx\,d\xi\right\}}\,\times\,\label{fa5}\\
\times\,\,\exp{\left\{u_0^2\int_0^t d\xi_1\int_0^{\xi_1}d\xi_2
\,\,[\,i{\bf k}\,+\,{\bf K}^*(a^*(\xi_1),a(\xi_1))\,]\cdot[\,i{\bf
k}\,+\,{\bf K}(a^*(\xi_2),a(\xi_2))\,]\,\right\}} \, \prod_{x,\,\xi}
\frac {da^*da}{2\pi i}\,\,\,,\,\label{fa5}
\end{array}
\end{equation}
where edge conditions\, $\,a^*(x,\xi=t)=0\,$, $\,a(x,\xi=0)=0\,$\,
must be satisfied,
\[
\begin{array}{l}
\widehat\Lambda_0\,=\, -\,{\bf v}\cdot\frac {\partial}{\partial
\rho}\,+\,\Phi^{\prime}(\rho)\cdot\frac {\partial}{\partial {\bf
p}}\,\,\label{L0}
\end{array}
\]
is Liouville operator of atom interacting with BP (fixed at
coordinate origin), and
\[
\begin{array}{l}
{\bf K}^*(a^*,a)\,=\,-\int c(x)\,\frac
{\Phi^{\prime}(\rho)}{T}\,\,a^*(x)\,dx\,+\,\int a^*(x)\left[\,\frac
{\partial}{\partial \rho}-\frac
{\Phi^{\prime}(\rho)}{2T}\right]\,a(x)\,dx\,\,\,,
\end{array}
\]
\[
\begin{array}{l}
{\bf K}(a^*,a)\,=\,\int c(x)\,\frac
{\Phi^{\prime}(\rho)}{T}\,\,a(x)\,dx\,+\,\int a^*(x)\left[\,\frac
{\partial}{\partial \rho}+\frac
{\Phi^{\prime}(\rho)}{2T}\right]\,a(x)\,dx\,\,\,,
\end{array}
\]
where\, $\,\int...\,dx=\int\int...\,d{\bf p}\,d\rho\,$\,.

Unfortunately, a correct non-perturbation treatment of such strongly
non-Gaussian path integrals as (\ref{fa3}) or (\ref{fa4}) or even
(\ref{fa5}) is in itself non-trivial problem \cite{sf}. To make use
of them, one has to construct some perturbation expansion (e.g. over
$\,m/M\,$ in the limit of hard BP).

\section{Generalized boson representation and continued fractions}

A different boson-like representation is prompted by characteristic
triple-diagonal form of the system of equations
(\ref{v01})-(\ref{vn1}). Considering $\,V_n\,$ as components of
vector in the Fock space, let us define in this space birth and
annihilation operators, $\,\widehat{\mathcal{A}}^{\,\dag}\,$ and
$\,\widehat{\mathcal{A}}\,$,\, $\,\widehat{\bf a}^{\,\dag}\,$ and
$\,\widehat{\bf a}\,$,\, as follows:
\[
\begin{array}{l}
\widehat{\mathcal{A}}^{\,\dag}\,\,V_n\,=\,\left({\bf V}+T\,\frac
{\partial }{\partial {\bf P}}\right )\cdot\left[\,G_m({\bf
p}_{n+1})\,E^{\,\prime}(\rho_{n+1})\,V_n\,+\,\sum_{j\,=1}^n G_m({\bf
p}_j)\,E^{\,\prime}(\rho_j)\, V_{n}(...\,x_j\rightarrow
x_{n+1}\,...)\right]\,\equiv\,\\
\,\,\,\,\,\,\,\,\,\,\,\,\,\,\,\,\,\,\,\equiv\,\left({\bf P}+TM\,\frac
{\partial }{\partial {\bf P}}\right )\cdot\,\widehat{\bf
a}^{\,\,\dag}\,V_n\,\,\,,
\end{array}
\]
\[
\begin{array}{l}
\widehat{\mathcal{A}}\,\,V_{n>\,0}\,=\,-\,\nu\, \frac {\partial
}{\partial {\bf P}}\int_{n} \Phi^{\prime}(\rho_{\,n})\,
V_{n}\,\equiv\,\frac {\partial }{\partial {\bf P}}\cdot\,\widehat{\bf
a}\,\,V_n\,\,\,,\,\,\,\,\,\,\,
\widehat{\mathcal{A}}\,\,V_0\,=\,0\,\,\,,\,\,\,\,\,\,\,\widehat{\bf
a}\,V_0\,=\,0\,\,\,,
\end{array}
\]
where symmetry of all the CF $\,V_n\,$ is taken in mind. According to
this definition,
\begin{equation}
\begin{array}{l}
(\,\widehat a_{\,\alpha}\,\widehat a^{\,\dag}_{\,\beta}\,- \,\widehat
a^{\,\dag}_{\,\beta}\,\widehat
a_{\,\alpha}\,)\,=\,a^2\,\delta_{\alpha\beta}\,\,\,,\\
(\,\widehat{\mathcal{A}}\widehat{\mathcal{A}}^{\,\dag}\,-
\,\widehat{\mathcal{A}}^{\,\dag}\widehat{\mathcal{A}}\,)\,V_n\,=\,
a^2\,\left[\,\frac {\partial }{\partial {\bf P}}\cdot\left({\bf
P}+TM\,\frac {\partial }{\partial {\bf P}}\right )\,+\,
\sum_{j\,=1}^n\,\widehat\Pi(j)\right]\,V_n\,\,\,,\label{comm}\\
a^2\,\equiv\,-\,\frac {\nu}{Md}\int \Phi^{\prime}(\rho)\cdot
E^{\prime}(\rho)\,d\rho\,\,\,,
\end{array}
\end{equation}
if we assume that $\,\Phi(\rho)\,$ is spherically symmetric potential
and define projection operator $\,\widehat\Pi\,$ as
\[
\begin{array}{l}
\widehat\Pi(j)\,V_n\,=\, d\,G_m({\bf
p}_j)\,E^{\,\prime}(\rho_j)\cdot\int \Phi^{\prime}(\rho_{j})
\,V_{n}\,d\rho_j\,d{\bf p}_j\,\left[\int \Phi^{\prime}(\rho)\cdot
E^{\prime}(\rho)\,d\rho\,\right]^{-1}\,\,
\end{array}
\]
Besides, introduce operator\, $\,\widehat{\Lambda}_{\,{\bf k}}\,$\,
by equalities\,\,\, $\,\widehat{\Lambda}_{\,{\bf
k}}\,V_0\,\equiv\,i({\bf k}\cdot{\bf V})\,V_0\,$\,\,\,,
\[
\widehat{\Lambda}_{\,{\bf k}}\,V_n\,\equiv\,\left[\,i({\bf
k}\cdot{\bf V})\,+\,\sum_{j\,=1}^n\widehat{\Lambda}(j)\,\right]V_n\,
\]
Then formal solution of equations (\ref{v01})-(\ref{vn1}), under
initial conditions (\ref{ibc}), in respect to Laplace transform of
$\,V_0(t,i{\bf k};\,\nu)\,$, can be represented by continued
fraction:
\[
\begin{array}{l}
\int_0^{\infty}e^{-\,zt}\,V_0(t,i{\bf k};\,\nu)\,dt\,=\,\\ =\,\int
d{\bf P}\,\left[\,z-\widehat{\Lambda}_{\,{\bf k}}-
\widehat{\mathcal{A}}\left[z-\widehat{\Lambda}_{\,{\bf
k}}-\widehat{\mathcal{A}} \,\left[\,z-\widehat{\Lambda}_{\,{\bf
k}}-...\,\right]^{-\,1} \widehat{\mathcal{A}}^{\,\dag}
\right]^{-\,1}\widehat{\mathcal{A}}^{\,\dag} \right]^{-\,1}G_M({\bf
P})\,=\,
\end{array}
\]
\begin{equation}
\begin{array}{l}
=\,\int d{\bf P}\,\left[\,z-i{\bf k}\cdot{\bf V}-\frac
{\partial}{\partial
P_{\alpha}}\,\,\widehat\Gamma^{(1)}_{\alpha\beta}\left(z,i{\bf k}
\right)\left(P_{\beta}+TM\frac {\partial}{\partial
P_{\beta}}\right)\right]^{-\,1}G_M({\bf P}) \,\,\,,\label{cfr}
\end{array}
\end{equation}
where $\,\widehat\Gamma^{(1)}_{\alpha\beta}\,$ begins recursive chain
of operators
\begin{equation}
\begin{array}{l}
\widehat\Gamma^{(n)}_{\alpha\beta}\left(z,i{\bf k}
\right)\,=\,\widehat a_{\,\alpha}\left[\,z-i{\bf k}\cdot{\bf
V}-\sum_{j\,=1}^n\widehat{\Lambda}(j)-\frac {\partial}{\partial
P_{\gamma}}\,\,\widehat\Gamma^{(n+1)}_{\gamma\delta}\left(z,i{\bf k}
\right)\left(P_{\delta}+TM\frac {\partial}{\partial
P_{\delta}}\right)\right]^{-\,1}\widehat a^{\,\dag}_{\,\beta}\,
\label{gn}
\end{array}
\end{equation}

\section{Stochastic form of the boson representation}
Let $\,\mathcal{A}^\dag\,$, $\,\mathcal{A}\,$ be a set of pairs of
boson birth and annihilation operators, such that
$\,\mathcal{A}_{\alpha}\mathcal{A}^\dag_{\beta}-
\mathcal{A}^\dag_{\beta}\mathcal{A}_{\alpha}=\delta_{\alpha\beta}\,$.
Then any path integral like (\ref{fa2}) or (\ref{fa3}) or (\ref{fa4})
or (\ref{fa5}),
\[
\begin{array}{l}
\int \exp{\left\{\int^t_0 \frac 12\,
(\,\dot{\mathcal{A}}^*\mathcal{A}\,-\,\mathcal{A}^*\dot{\mathcal{A}}\,)
\,d\xi\,+\,\mathbb{F}\{\mathcal{A}^*(\xi),\mathcal{A}(\xi)\}\right\}}
\,\prod_{\xi} \frac {d\mathcal{A}^*d\mathcal{A}}{2\pi
i}\,\equiv\,\\
\,\,\,\,\,\,\,\,\,\,\,\,\,\,\,\,\,\,\,\,\,\,\,\,\,\equiv\,\int
\exp{\left[\,\,\mathbb{F}\{\mathcal{A}^*(\xi),\mathcal{A}(\xi)\}\,\,\right]}\,
\,d\mathcal{M}\{\mathcal{A}^*,\mathcal{A}\} \,\,\,,
\end{array}
\]
with arbitrary functional
$\,\exp{\left[\,\mathbb{F}\{\mathcal{A}^*(\xi),\mathcal{A}(\xi)\}\,\right]}\,$
defined on interval $\,0\leq\xi\leq t\,$,
 can be formally considered as averaging of this functional over Gaussian
 ``probabilistic'' measure $\,d\mathcal{M}\{\mathcal{A}^*,\mathcal{A}\} \,$\,:
\begin{equation}
\begin{array}{l}
\int
\exp{\left[\,\,\mathbb{F}\{\mathcal{A}^*(\xi),\mathcal{A}(\xi)\}\,\,\right]}\,
\,d\mathcal{M}\{\mathcal{A}^*,\mathcal{A}\} \,=\,\left\langle\,
\exp{\left[\,\,\mathbb{F}\{\mathcal{A}^*(\xi),\mathcal{A}(\xi)\}\,\,\right]}
\,\right\rangle\,\,\label{sbr}
\end{array}
\end{equation}
The measure\, $\,d\mathcal{M}\{\mathcal{A}^*,\mathcal{A}\} \,$\, is
completely characterized by corresponding pair correlation functions:
\begin{equation}
\begin{array}{l}
\langle\,\mathcal{A}_{\alpha}(t_1)\,\mathcal{A}_{\beta}(t_2)\,\rangle\,=
\,\langle\,\mathcal{A}^*_{\alpha}(t_1)\,\mathcal{A}^*_{\beta}(t_2)\,
\rangle\,=\,0\,\,\,,
\,\,\,\,\,\,\langle\,\mathcal{A}_{\alpha}(t_1)\,
\mathcal{A}^*_{\beta}(t_2)\,\rangle\,
=\,\delta_{\alpha\beta}\,\Theta(t_1-t_2)\,\,\,,\label{sm}
\end{array}
\end{equation}
where $\,\Theta(t)\,$ is Heaviside step function. To prove these
equalities, it is sufficient to calculate Gaussian integral which
represents characteristic functional of the stochastic processes
$\,\mathcal{A}^*(t)\,$ and $\,\mathcal{A}(t)\,$\,,
\begin{equation}
\begin{array}{l}
\int \exp{\left\{\,\int_0^t
[\,b(\xi)\cdot\mathcal{A}^*(\xi)+b^*(\xi)\cdot\mathcal{A}(\xi)\,]\,d\xi
\,\right\}}\, d\mathcal{M}\{\mathcal{A}^*,\mathcal{A}\} \,=\,
\exp{\left[\,\int_0^t dt^{\,\prime}
\int_0^{t^{\,\prime}}dt^{\,\prime\prime}\,\,b^*(t^{\,\prime})\cdot
b(t^{\,\prime\prime})\,\right]}\,\,\,,\label{cfa}
\end{array}
\end{equation}
where the mentioned edge conditions are taken in mind, and
$\,b(t)\,$, $\,b^*(t)\,$ are arbitrary time functions.

In this sense, in particular,
\begin{equation}
\begin{array}{l}
V_0(t,i{\bf k};\,\nu)\,=\,\langle
0|\,\,e^{\,t\pounds(\mathcal{A}^\dag,\,\mathcal{A})}\,
\,|0\rangle\,=\,\left\langle \exp{\int^t_0
\pounds(\mathcal{A}^*(\xi),\mathcal{A}(\xi))\, d\xi}\right\rangle
\,\,\label{fa6}
\end{array}
\end{equation}
with operator function\,
$\,\pounds(\mathcal{A}^\dag,\,\mathcal{A})\,$ defined by
(\ref{mop1})-(\ref{mop3}), and
\begin{equation}
\begin{array}{l}
V_0(t,i{\bf k};\nu)=\left\langle \exp{\left[iu_0{\bf k}\cdot\int^t_0
\left[{\bf A}(\xi)+{\bf A}^*(\xi)
\right]d\xi-\int_{t>\,\xi_1>}\int_{\xi_2>\,0} {\bf A}^*(\xi_1)
\cdot\mathbb{G}\{\xi_1,\xi_2,{\bf A}^*,{\bf A}\}\cdot {\bf
A}(\xi_2)\,\right]}\right\rangle\label{sv}
\end{array}
\end{equation}
instead of (\ref{fa3}), with Gaussian random processes $\,{\bf
A}^*(t)\,$ and $\,{\bf A}(t)\,$ defined by
\begin{equation}
\begin{array}{l}
\langle\,{\bf A}_{\alpha}(t_1)\,{\bf A}_{\beta}(t_2)\,\rangle\,=
\,\langle\,{\bf A}^*_{\alpha}(t_1)\,{\bf A}^*_{\beta}(t_2)\,
\rangle\,=\,0\,\,\,, \,\,\,\,\,\,\,\,\,\,\,\,\langle\,{\bf
A}_{\alpha}(t_1)\, {\bf A}^*_{\beta}(t_2)\,\rangle\,
=\,\delta_{\alpha\beta}\,\Theta(t_1-t_2)\,\,\label{sm1}
\end{array}
\end{equation}
Such ``stochastic'' point of sight at calculation of path integrals
can make it more constructive.

Notice that for any functional
$\,\mathbb{F}\,=\,\mathbb{F}\{\mathcal{A}^*(\xi),\mathcal{A}(\xi)\}\,$
arranged like second term in (\ref{sv}),
\begin{equation}
\mathbb{F}\{\mathcal{A}^*(\xi),\mathcal{A}(\xi)\}\,=\,
-\int_{t>\,\xi_1>}\int_{\xi_2>\,0} \mathcal{A}_{\alpha}^*(\xi_1)
\,\,\mathbb{G}_{\alpha\beta}\{\xi_1,\xi_2,\,\mathcal{A}^*,\mathcal{A}\}\,\,
\mathcal{A}_{\beta}(\xi_2)\,\,d\xi_2\,d\xi_1\,\,\,,\label{sf1}
\end{equation}
where\,
$\,\mathbb{G}_{\alpha\beta}\{\xi_1,\xi_2,\,\mathcal{A}^*,\mathcal{A}\}\,$\,
involves $\,\mathcal{A}^*(\xi)\,$ and $\,\mathcal{A}(\xi)\,$ from
interval\, $\,\xi_2\leq\xi\leq\xi_1\,$\, only, the identities
\[
\begin{array}{l}
\left\langle\,\mathbb{F}^{\,n}\{\mathcal{A}^*(\xi),
\mathcal{A}(\xi)\}\,\right\rangle\,=\,0\,\,\,\,\,
\,\,(n>0)\,\,\,,\,\,\,\,\,\,\,\,\,\,
\left\langle\,\exp{\left[\,\mathbb{F}\{\mathcal{A}^*(\xi),
\mathcal{A}(\xi)\}\,\right]}\,\right\rangle\,=\,1\,\,
\end{array}
\]
take place, that is by itself such functional is identical to zero.
In special case of quadratic (bilinear) functional, when\,
$\,\mathbb{G}_{\alpha\beta}\{\xi_1,\xi_2,\,\mathcal{A}^*,\mathcal{A}\}
=\mathcal{G}_{\alpha\beta}(\xi_1-\xi_2)\,$,
\begin{equation}
\begin{array}{l}
\left\langle\,\exp{\left(\,\int_0^t
[\,b(\xi)\cdot\mathcal{A}^*(\xi)+b^*(\xi)\cdot\mathcal{A}(\xi)\,]\,d\xi
\,-\int_{t>\,\xi_1>}\int_{\xi_2>\,0} \mathcal{A}_{\alpha}^*(\xi_1)
\,\,\mathcal{G}_{\alpha\beta}(\xi_1-\xi_2)\,\mathcal{A}_{\beta}(\xi_2)
\,\right )}\,
\right\rangle\,=\,\\
\,\,\,\,\,\,\,\,\,\,\,\,\,\,\,\,\,\,=\,\,\exp{\left\{\,\int_0^t
dt^{\,\prime}
\int_0^{t^{\,\prime}}dt^{\,\prime\prime}\,\,\,b^*_{\alpha}(t^{\,\prime})\,
Q_{\alpha\beta}(t^{\,\prime}-t^{\,\prime\prime})\,
b_{\beta}(t^{\,\prime\prime}) \,\right\}}\,\,\,,\label{sf20}
\end{array}
\end{equation}
where matrix function\, $\,Q_{\alpha\beta}(t)\,$\, is defined by
\begin{equation}
\begin{array}{l}
\,\,\,\,Q\,\equiv\,\Theta\,[\,1\,+\,\otimes\,\,\mathcal{G}\,
\otimes\,\Theta\,]^{-1}\,\,\,,\,\,\,\,\,\,\,\int_0^\infty
e^{-zt}\,Q(t)\,dt\,=\,\left[\,z\,+\,\int_0^\infty
e^{-zt}\,\mathcal{G}(t)\,dt\,\right]^{-\,1}\,\,\,,\label{conv}
\end{array}
\end{equation}
with symbol\, $\,\otimes\,$ standing for time convolution. The first
of the latter expressions remains valid also when
$\,\mathcal{G}_{\alpha\beta}=\mathcal{G}_{\alpha\beta}(\xi_1,\xi_2)\,$
is not a difference kernel.

\section{Discussion and conclusion}
At present, unfortunately, none of formally exact expressions
(\ref{fa}), (\ref{fa1}), (\ref{fa3}), (\ref{fa4}), (\ref{fa5}),
(\ref{cfr}) or (\ref{sv}) can be calculated exactly or at least
correctly, at least in the long-time limit
$\,t/\tau_p\rightarrow\infty\,$ with $\,\tau_p\,$ denoting relaxation
time of BP's momentum. Therefore it remains only to discuss the
convenient approximation of exact theory and establish discreditable
invalidity of this approximation from the point of view of exact
``virial relations'' (\ref{vexpn})-(\ref{inf0}) (see
\cite{last,may,mpa2,pro,mpa3,jstat} and remark at end of Sec.2).

{\bf 1\,}.\, The mentioned approximation follows from the chain
(\ref{v01})-(\ref{vn1}) if we cut off it already at second level,
i.e. neglect second-order (three-particle) correlation and thus all
higher-order correlations. This is just what one always makes
(knowingly or unknowingly) when creating Boltzmannian kinetics. Then
\begin{equation}
V_1(t,i{\bf k},{\bf P},\rho,{\bf p};\,\nu)\,=\,\int_0^t e^{[\,i\,{\bf
k}\cdot{\bf
V}\,+\,\widehat{\Lambda}\,]\,(t-\,t^{\,\prime})}\,G_m({\bf
p})\,E^{\,\prime}(\rho)\cdot\left({\bf V}+T\,\frac {\partial
}{\partial {\bf P}}\right )V_0(t^{\,\prime},i{\bf k},{\bf
P};\,\nu)\,dt^{\,\prime}\,\,\,,\label{v1a}
\end{equation}
and the first BBGKY equation (\ref{v01}) turns into closed kinetic
equation
\begin{equation}
\frac {\partial V_0(t,i{\bf k},{\bf P};\nu)}{\partial t}\,=\,i({\bf
k}\cdot{\bf V})\,V_0(t)\,+\,\frac {\partial }{\partial
P_{\alpha}}\int_0^t
\widehat{\mathcal{G}}_{\alpha\beta}(t-t^{\,\prime},i{\bf
k})\,\left(P_{\beta}+TM\,\frac {\partial }{\partial P_{\beta}}\right
)V_0(t^{\,\prime})\,\,dt^{\,\prime}\,\,\,,\label{v0a}
\end{equation}
with $\,V_0(t)\equiv V_0(t,i{\bf k},{\bf P};\nu)\,$ on the right and
operator-valued kernel
\begin{equation}
\widehat{\mathcal{G}}_{\alpha\beta}(\theta,i{\bf k})\,\equiv\,\frac
{\nu}{TM}\int d\rho\int d{\bf p}\,\,\,
\Phi^{\prime}_{\alpha}(\rho)\,\,e^{[\,i\,{\bf k}\cdot{\bf
V}\,+\,\widehat{\Lambda}\,\,]\,\theta\,}\,
\Phi^{\prime}_{\beta}(\rho)\,G_m({\bf p})\,E(\rho)\,\,\label{kern0}
\end{equation}
From the point of view of continued fraction (\ref{cfr}), this is
``one-loop approximation'' when one substitutes zero for
$\,\widehat\Gamma^{(2)}_{\alpha\beta}\,$ and thus leaves two floors
of the fraction only.

Evidently, this kernel is a sharp function of $\,\theta\,$, with
width nearly equal to time duration of BP-atom collision,
$\,\tau_0=r_0/v_0\,$. At that, principally we are interested in the
long-time long-range limit only, when $\,t/\tau_p
\rightarrow\infty\,$ and $\,{\bf k}\rightarrow 0\,$ under $\,{\bf
k}^2 t=\,$const\,. Therefore factor $\,i({\bf k}\cdot{\bf V})\,$ in
(\ref{kern0}) can be neglected, and equation (\ref{v0a}) reduces to
the ``Boltzmann-Lorentz equation''
\begin{equation}
\begin{array}{l}
\frac {\partial V_0}{\partial t}\,=\,i({\bf k}\cdot{\bf
V})\,V_0\,+\,\widehat{\mathcal{B}}\, \,V_0\,\,\,,
\,\,\,\,\,\,\,\,\widehat{\mathcal{B}}\,\equiv\, \frac {\partial
}{\partial {\bf P}}\cdot\int_0^\infty
\widehat{\mathcal{G}}(\theta,0)\,d\theta\,\cdot \left({\bf P}+TM\,
\frac {\partial }{\partial {\bf P}}\right )\,\,\,,\label{v0b}
\end{array}
\end{equation}
where $\,\widehat{\mathcal{B}}\,$ plays role of linearized collision
operator.

Interestingly, above ``derivation'' of Boltzmannian kinetics had not
required the Boltzmann's ``Stosszahl-ansatz''. Though the latter is
necessary if one wants to transform $\,\widehat{\mathcal{B}}\,$ into
standard ``collision integral''. This observation shows that the
heart of Boltzmannian kinetics is neglect of third- and higher-order
correlations (and thus, in essence, neglect of true second-order
correlation).

Recall that $\,\widehat{\Lambda}\,$ contains $\,{\bf P}\,$ and
$\,\partial/\partial {\bf P}\,$, therefore in general operator
$\,\widehat{\mathcal{G}}(\theta,0)\,$ significantly depends on
$\,{\bf P}\,$ and $\,\partial/\partial {\bf P}\,$. But under the
long-range limit it effectively interchanges to $\,\int d{\bf P}\,
\widehat{\mathcal{G}}(\theta,0)\,G_M({\bf P})\,$. Therefore,
regardless of details of $\,\widehat{\mathcal{B}}\,$, long-range
asymptotic what follows from (\ref{v0a}) or (\ref{v0b}) is the
Gaussian one:
\begin{equation}
\begin{array}{l}
V_0\left(t,i{\bf k},{\bf P}; \nu\right)\,\rightarrow\,
\exp{\left[-D(\nu)\,{\bf k}^2 t\,\right]}\,G_M({\bf P})\left(1+\frac
{\tau_p}{M}\,\,i{\bf k}\cdot{\bf
P}+...\right)\,\,\,,\\
V_0\left(t,i{\bf k};\nu\right)\,\rightarrow\,\exp{[-D(\nu)\,{\bf k}^2
t\,]}\,\,\,,\label{gs}
\end{array}
\end{equation}
where BP's diffusivity $\,D(\nu)\,$ and momentum relaxation time are
presented by
\begin{equation}
\begin{array}{l}
D(\nu)\,=\,u_0^2\,\tau_p\,\propto\,\nu^{-1}\,\,\,,\\
\frac {1}{\tau_p}\,\equiv\,\frac {\nu}{TMd}\,\int_0^\infty
d\theta\int d{\bf P} \int d\rho\int d{\bf
p}\,\,\,\Phi^{\prime}(\rho)\cdot
e^{\,\widehat{\Lambda}\,\theta}\,\Phi^{\prime}(\rho)\,\,G_m({\bf
p})\,G_M({\bf P})\,E(\rho)\,=\,\,\label{d}\\
\,\,\,\,\,\,\,\,\,=\,\frac {2m}{M+m}\,\,\nu\,\int\int \left|{\bf v} -
{\bf V}\right|\,\Sigma(|{\bf v} - {\bf V}|)\,\,G_m({\bf p})\,G_M({\bf
P})\,d{\bf p}\,d{\bf P}\,
\end{array}
\end{equation}
In (\ref{gs}) last multiplier of $\,V_0\left(t,i{\bf k},{\bf P};
\nu\right)\,$ is important for (\ref{v1a}), while the dots replace
unimportant terms, and in (\ref{d}) $\,\Sigma\,$ is full effective
cross-section of BP-atom collisions.

All that seems beautiful till one confronts equations (\ref{v1a}) and
(\ref{v0b}) to the simplest virial relation (\ref{inf0}). The latter
requires that
\begin{eqnarray}
\begin{array}{l}
\frac {\partial V_0(t,i{\bf k};\,\nu)}{\partial \nu}\,=\,\int\int\int
V_1(t,i{\bf k},{\bf P},\rho,{\bf p};\,\nu)\,\, d{\bf p}\,d\rho\,d{\bf
P}\,\,\,\label{er}
\end{array}
\end{eqnarray}
Combining this identity from exact theory with (\ref{gs}) and
(\ref{d}) we see that if the mentioned approximation was correct then
we would have
\begin{equation}
\begin{array}{l}
\int\int\int V_1(t,i{\bf k},{\bf P},\rho,{\bf p};\,\nu)\,\, d{\bf
p}\,d\rho\,d{\bf P}\,\rightarrow\, \frac {D(\nu)\,{\bf
k}^2\,t}{\nu}\,e^{-\,D(\nu)\,{\bf k}^2\,t}\,\,\label{gs1}
\end{array}
\end{equation}
In fact, however, expression (\ref{v1a}), as combined with (\ref{gs})
and (\ref{d}), after quite standard (although rather troublesome)
manipulations yields
\begin{equation}
\begin{array}{l}
\int\int\int V_1(t,i{\bf k},{\bf P},\rho,{\bf p};\,\nu)\,\, d{\bf
p}\,d\rho\,d{\bf P}\,\rightarrow\,\\
\,\,\,\,\,\,\,\,\,\,\,\,\,\,\,\,\,\,\rightarrow\,-\,\frac
{u_0^2\,{\bf k}^2}{\nu} \int_0^t \exp{\left[\,-\frac 12\,u_0^2\,{\bf
k}^2\,(t-t^{\,\prime})^2\,-\,D(\nu)\,{\bf
k}^2\,t^{\,\prime}\,\right]}\,(t-t^{\,\prime})\,
\,dt^{\,\prime}\,\rightarrow\,-\,\frac {1}{\nu}\,\,e^{-\,D(\nu)\,{\bf
k}^2\,t}\,\,\,,\label{v3a}
\end{array}
\end{equation}
since exponential in (\ref{v1a}) corresponds to free ballistic flight
of BP after its single collision with an atom.

The difference between (\ref{gs1}) and (\ref{v3a}) is qualitative,
and it says that the conventional picture, including the
Boltzmann-Lorentz equation, is far from truth!

Such strong discrepancy came from our neglect of the third-order
correlations in (\ref{v11}), i.e. $\,V_2$'s contribution to
(\ref{v11}), and thus neglect of all higher-order correlations.
However, the ``virial expansions'' (\ref{vexpn}), in particular,
first of them, as written via Fourier transforms (\ref{nf}),
\begin{eqnarray}
\begin{array}{l}
V_0(t,i{\bf k},{\bf P};\,\nu+\sigma \nu )\,=\,V_0(t,i{\bf k},{\bf
P};\,\nu)\,+\sum_{n\,=\,1}^\infty \frac
{(\nu\sigma)^{\,n}}{n!}\int_{1} ...\int_{n} V_{n}(t,i{\bf
k},\rho^{(n)},{\bf P},{\bf p}^{(n)};\,\nu)\,\,\label{vexp0}
\end{array}
\end{eqnarray}
(recall that $\,\sigma >-1\,$), help us to understand that cutting of
any even high correlations is bad idea.

Indeed, gas densities on two sides of (\ref{vexp0}), $\,\,\nu+\sigma
\nu \,$ and $\,\nu\,$, can be different in arbitrary strong
proportion $\,0<\sigma +1<\infty\,$. Therefore any cutoff in the
infinite series in (\ref{vexp0}) would give a faulty result (like
e.g. cutoff in the series representing exponential function). Hence,
in practice all terms on right-hand side of (\ref{vexp0}) are equally
important, regardless of value of gas density!

Consequently, all equations of the hierarchy (\ref{v0})-(\ref{vn}) or
(\ref{v01})-(\ref{vn1}) are equally important for correct analysis of
the BP's path probability distribution, even in the ``low density
limit'' (``Boltzmann-Grad limit''). This means that Boltzmannian
kinetics is not a true ``zero-order approximation'' of rigorous
theory in respect to the density. Or, better to say, true kinetics
has no literal ``zero-order approximation'' at all.

{\bf 2\,}.\, The reason for all this was explained more than once in
\cite{bk12,bk3} twenty five years ago, then in \cite{i1} and later in
\cite{i2,p1,p2,i3,kmg,kmg1} and \cite{mpa2,pro,mpa3,jstat,jsp} (and
in principal sense anticipated in \cite{kr} sixty years ago).
Indifference of many-particle system to a number of happened events
of definite kind (BP's collisions with gas atoms here) leads to
scaleless fluctuations (``1/f\,-noise'') in ``mean number of events
per unit time'' and related quantities (e.g. BP's diffusivity and
mobility here).

In spite of this understanding, one can envy creators of conventional
kinetics (see Introduction): it already resolved all its problems.
But, from the other hand, this is kinetics of an invented tiresome
and ``dead'' world. Our consideration demonstrated that even such
simple world as Brownian particle interacting with ideal gas
apparently is ``alive'' and interesting. Nobody is able to predict
what ``number of collisions per unit time'' will meet this particle
in particular life. Hence, real theory is not in the past, it is yet
in the future.

\end{document}